\documentclass[a4paper,british,twocolumn, showpacs]{revtex4}
\usepackage[T1]{fontenc}
\usepackage[latin9]{inputenc}
\usepackage{graphicx}
\usepackage{amssymb}

\usepackage{times}

\usepackage{babel}

\begin{document}

\title{A Non-Demolition Single Spin Meter}
\begin{abstract}
We present the theory of a single spin meter consisting of a quantum
dot in a magnetic field under microwave irradiation combined with
a charge counter. We show that when a current is passed through the
dot, a change in the average occupation number occurs if the microwaves
are resonant with the on-dot Zeeman splitting. The width of the resonant
change is given by the microwave induced Rabi frequency, making the
quantum dot a sensitive probe of the local magnetic field and enabling
the detection of the state of a nearby spin. If the dot-spin and the
nearby spin have different g-factors a non-demolition readout of the
spin state can be achieved. The conditions for a reliable spin readout
are found.
\end{abstract}

\author{J. Wabnig}

\affiliation{Department of Materials, Oxford University, Oxford OX1 3PH, United
Kingdom}

\author{B. W. Lovett}

\affiliation{Department of Materials, Oxford University, Oxford OX1 3PH, United
Kingdom}


\maketitle
Single-shot readout of individual spins lies at the heart of many
future technologies, from spintronics and quantum computation to single
molecule spin resonance. There have been several recent proposals
and demonstrations of single spin read out, all with restricted applicability,
relying on the details of the physical system in which the spin is
housed \citep{2008Natur.453.1043H} or requiring certain special features,
such as optical activity \citep{maze:644}, nuclear spins \citep{2007arXiv0711.2343S}
or a large detector-system interaction \citep{2008PhRvB..77w5309L};
others are too slow to achieve a measurement in a single attempt \citep{2004Natur.430..329R}.
Here we introduce a versatile spin meter that can achieve reliable
determination of the spin state. Our electrical method is non-invasive,
in contrast to standard techniques that destroy the spin in the course
of the measurement \citep{2004Natur.430..431E}. It consists of a
quantum dot under microwave illumination and can in principle be used
to detect arbitrarily weakly coupled spins in one shot. 

Our scheme could be used to read out an arbitrary spin, and we will
focus our discussion on a system that cannot be measured by any means
proposed so far: a molecular spin that is not optically active. Single
molecule magnets exhibit a range of quantum phenomena \citep{2008NatMa...7..179B}
and represent a possible implementation of spin qubits \citep{2007PhRvL..98e7201A}.
One of the leading candidates is a fullerene with an endohedral nitrogen
(N@C$_{60}$), whose electron spin has a remarkably long coherence
time \citep{morton:014508}. However, the very property that makes
the spin coherence time so long, i.e. weak interaction of the spin
with the environment, also makes the readout difficult. One path to
single spin readout is to place the molecule in a tiny gap between
metal contacts; recent electrical current measurements through an
endohedral fullerene show spin dependent features in the current voltage
characteristic of the device \citep{2008arXiv0805.2585G}. The problem
with this approach is the unpredictable tunnel coupling of the molecule
to the metal leads. In this Letter, we shall show that this difficulty
can be overcome by spatially separating the electrical readout from
the molecular spin that is measured. 

\begin{figure}
\begin{centering}
\includegraphics[width=1\columnwidth]{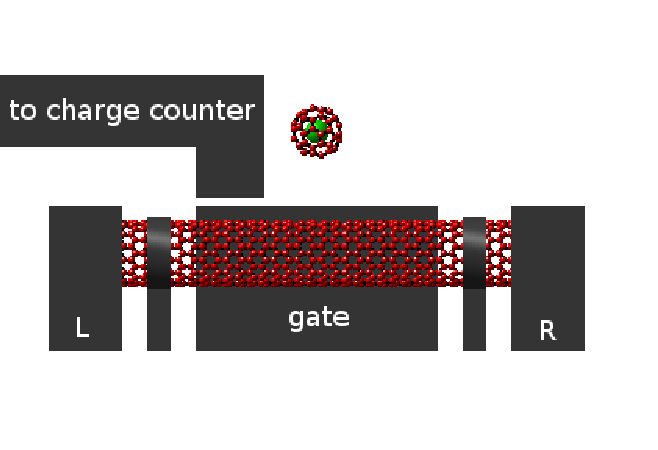}
\par\end{centering}

\caption{\label{fig1}An implementation of the spin meter concept. The central
part of the device is a quantum dot that is defined by electrical
gates which pinch off a one-dimensional conductor, such as a nanowire
or a carbon nanotube. The electrical transport through the quantum
dot can be controlled with high precision by varying the source and
drain potentials and the potentials of the extra finger gates placed
at the junctions. The charging state of the dot can be sensed through
an electrometer charge counter which is coupled to the quantum dot
by a conducting bridge, and a back gate can control the number of
electrons on the quantum dot. A single spin, e. g. in a fullerene,
is placed close to the quantum dot. If there is no electron hopping
between the dot and the fullerene, then the main coupling mechanism
between the electron spin on the dot and the fullerene spin is the
dipole-dipole interaction; there may also be an exchange interaction
that could be tuned by inserting a conducting molecule between the
fullerene and the nanotube. In such a setup no current will flow directly
through the fullerene and thus the spin will not be disturbed. }

\end{figure}

The set-up for our proposed spin meter is shown schematically in Fig.
\ref{fig1}. %
 A general obstacle in the sort of device we are proposing is the
wide range of relevant energy scales, which at first seem to preclude
sensitive spin measurement. An estimate of the dipole-dipole interaction
energy between two electron spins is $J\approx100\, r/r_{0}\,$MHz,
where $r_{0}=1$ nm. For a successful measurement, the current flow
through the dot has to be sensitive to a corresponding energy change.
The energy resolution of the dot is usually governed by the width
of a typical electronic energy level, which depends on the tunnelling
rate and, more importantly, by the temperature dependent width of
the Fermi function for the electrons in the leads. With attainable
temperatures in standard dilution refrigerators ($\sim10\,$ mK) the
energy resolution of the dot becomes $\sim1$ GHz, clearly not enough
to resolve the dipole-dipole interaction. 

This obstacle can be overcome by considering spin resonance effects
that are introduced by applying microwave radiation. Recently there
has been an increasing interest in methods for electrical detection
of spin resonance \citep{2006NatPh...2..835S,2004Natur.430..435X}.
Mozyrksy and Martin showed that the current in a quantum channel next
to an impurity spin depends on frequency of the incident microwave
frequency \citep{2003PhRvL..90a8301M}. We showed that a similar resonant
change in the current takes place in transport through a quantum dot
\citep{2008arXiv0804.0771W}. The line-width of the resonance introduces
a new energy scale into the energy dependence of the dot occupations
that can be much smaller than the temperature, subsequently enabling
the detection of much smaller shifts in energy. Typical microwave
intensities correspond to frequencies $\sim1\,\textnormal{MHz}$ but
can be made smaller, provided that the lifetime of the electron on
the dot is sufficient to allow for equalisation of the spin populations
on resonance, i. e. the tunnelling rate should be smaller than the
microwave induced Rabi frequency. Given a controllable tunnelling
rate the sensitivity of the dot detector can be increased by lowering
the microwave intensity and also the tunnelling rate, with the lower
limit set by the coherence time of the spin on the dot. 

The simplest model capturing all the relevant physics consists of
two spin 1/2 particles, one situated on the quantum dot (the flying
spin), the other nearby (the stationary spin). In a magnetic field
the energy levels for spin up and spin down for each of the spins
will be split by the Zeeman energy, but in general the splitting will
be different for the flying and the stationary spin whose g-factors
are typically unequal. Thus any incident narrow-band microwave radiation
can only be resonant with one of the spins. We choose it to be resonant
or nearly resonant with the flying spin. The dipole-dipole and exchange
coupling between the two spins give rise to an effective Ising interaction,
since flip-flop processes are suppressed by the Zeeman terms. The
resulting Hamiltonian in the rotating frame after the rotating wave
approximation can be written as\[
H=\frac{\Delta_{0S}}{2}\sigma_{z}^{(S)}+\frac{\Delta_{0F}}{2}\sigma_{z}^{(F)}+\frac{\Omega}{2}\sigma_{x}^{(F)}+\frac{J}{2}\sigma_{z}^{(S)}\sigma_{z}^{(F)},\]
where S labels the stationary spin and F the flying spin, $\Delta_{0S/F}=g_{S/F}\mu_{B}B_{z}-\hbar\omega_{0}$,
with $B_{z}$ the magnetic field in the z-direction, $\mu_{B}$ the
Bohr magneton, $g_{S/F}$ the g-factor for the stationary/flying spin
respectively and $\omega_{0}$ the microwave frequency. The Rabi frequency
due to an oscillating magnetic field in the x-direction for the flying
spin is $\Omega$; we can neglect the equivalent term for the off-resonant
static spin. $J$ is the Ising interaction strength. The Hamiltonian
separates into two decoupled subspaces, associated with stationary
spin up ($\uparrow$) and spin down ($\downarrow$) respectively,
and is written as follows:\[
H_{\uparrow/\downarrow}=\frac{1}{2}\Delta_{0\uparrow/\downarrow}\sigma_{z}^{(F)}+\frac{\Omega}{2}\sigma_{x}^{(F)}\mp\frac{\Delta_{0S}}{2}\]
where we defined the spin subspace dependent detuning $\Delta_{0\uparrow/\downarrow}=\Delta_{0F}\mp J$.
The upper/lower sign refers to the up/down subspace.

\begin{figure}
\begin{centering}
\includegraphics[width=1\columnwidth]{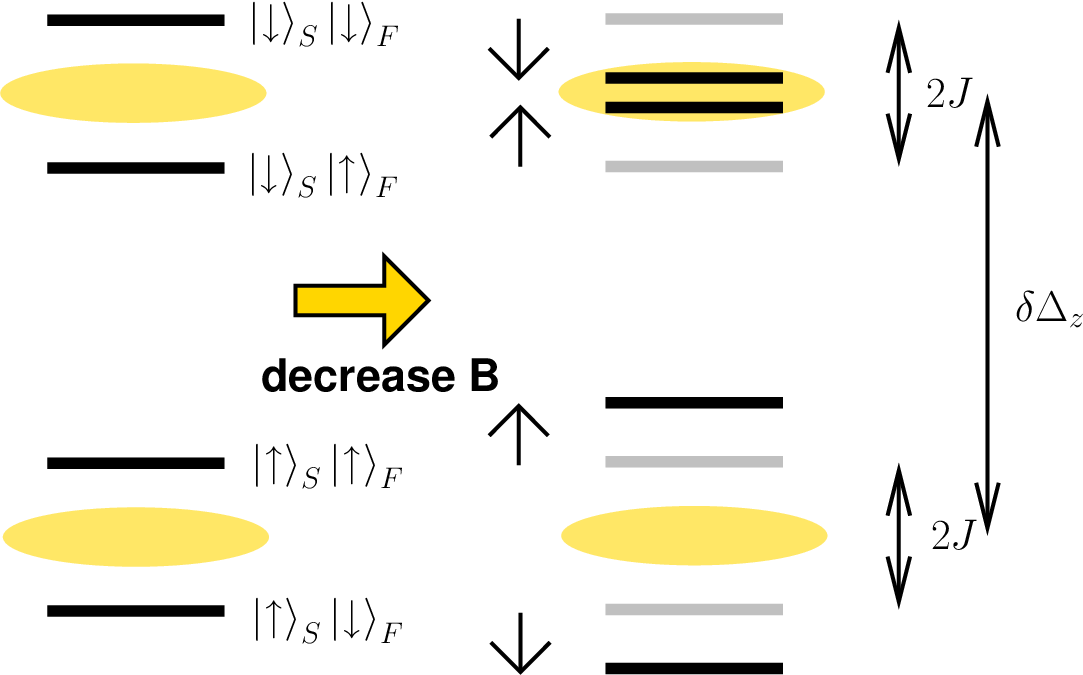}
\par\end{centering}

\caption{\label{fig2}The level structure of the two spin system in the rotating
frame for $\Delta_{0F}=0$. After decreasing the magnetic field the
upper two levels (stationary spin down subspace) move into resonance.
The difference in the Zeeman splittings is given by $\delta\Delta_{z}=\Delta_{0S}-\Delta_{0F}$.}

\end{figure}

The resulting energy level structure is shown in Fig.~\ref{fig2}.
Decreasing the magnetic field will shift the upper pair of levels
towards resonance, with the Rabi frequency dominating their splitting.
Meanwhile, the splitting of the lower pair increases. Close to resonance
the eigenstates become products of the static spin state with linear
superpositions of flying spin up and flying spin down:\begin{eqnarray*}
\left|a\right\rangle _{F} & = & a_{s\uparrow}\left|\uparrow\right\rangle _{F}+a_{s\downarrow}\left|\downarrow\right\rangle _{F}\\
\left|b\right\rangle _{F} & = & b_{s\uparrow}\left|\uparrow\right\rangle _{F}+b_{s\downarrow}\left|\downarrow\right\rangle _{F},\end{eqnarray*}
$s=\uparrow,\,\downarrow$ denotes the stationary spin subspace. The
detuning dependent coefficients are given by\[
a_{s\uparrow/\downarrow}=\sqrt{\frac{1}{2}\left(1\pm\frac{\Delta_{0s}}{\sqrt{\Delta_{0s}^{2}+\Omega^{2}}}\right)},\]
where $b_{s\uparrow}=-a_{s\downarrow}$ and $b_{s\downarrow}=a_{s\uparrow}$.

\begin{figure}
\begin{centering}
\includegraphics[width=1\columnwidth]{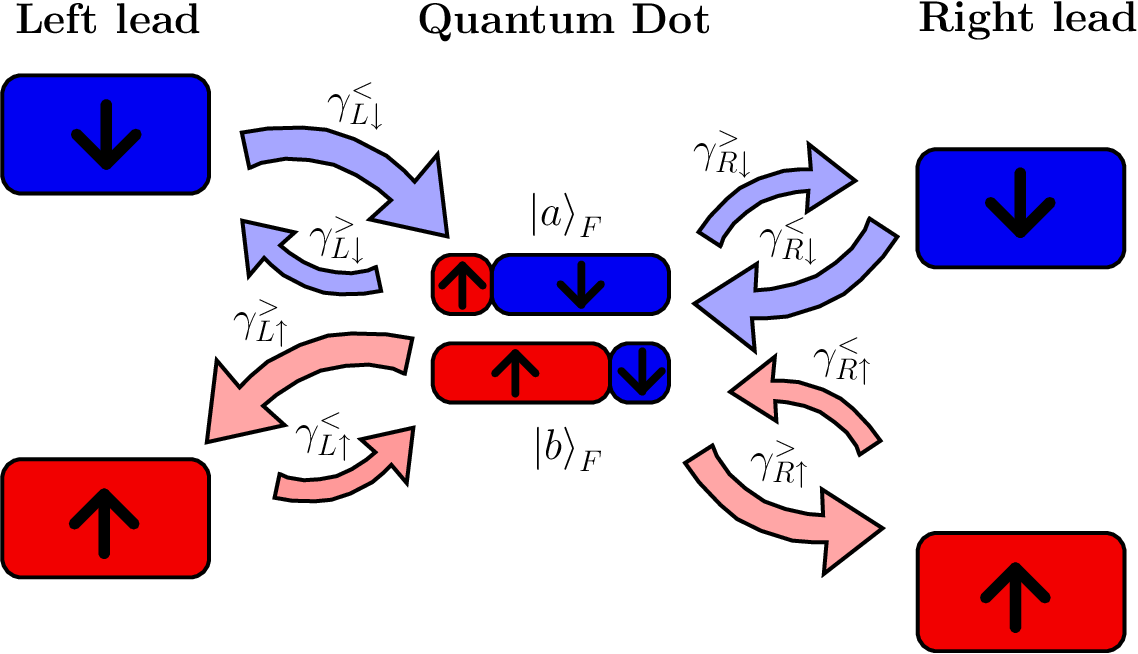}
\par\end{centering}

\caption{\label{fig3}A schematic diagram showing tunnelling off and onto the
quantum dot. The states on the quantum dot are superpositions of spin
up and spin down. The rates for tunnelling on and off the quantum
dot depend on the spin, the bias voltage, temperature, Zeeman splitting
and gate voltage.}

\end{figure}

Having established a model we can now ask whether we retrieve information
about the interacting two spin system by passing a current through
the quantum dot while monitoring the average population on the dot.
Considering only the stationary spin up subspace (or similarly the
spin down subspace) we can analyse the rates of tunnelling from the
leads onto the dot and from the dot into the leads. A schematic depiction
of the tunnelling rates is shown in Fig. \ref{fig3}. 

We can derive the rate equations for the stationary spin up subspace,
as\begin{eqnarray}
\dot{p}_{\uparrow a} & = & \left(\left|a_{\downarrow}\right|^{2}\gamma_{\downarrow}^{<}+\left|a_{\uparrow}\right|^{2}\gamma_{\uparrow}^{<}\right)p_{\uparrow0}\nonumber \\
 &  & -\left(\left|a_{\downarrow}\right|^{2}\gamma_{\downarrow}^{>}+\left|a_{\uparrow}\right|^{2}\gamma_{\uparrow}^{>}\right)p_{\uparrow a},\label{pa}\end{eqnarray}
where $p_{\uparrow a}$ is the probability of finding the stationary
spin in state $\uparrow$ and the flying spin in state $|a\rangle_{F}$
and $p_{\uparrow0}$ is the probability of finding the stationary
spin in state $\uparrow$ and no electron on the dot, $\gamma_{\uparrow/\downarrow}^{<}$
are the tunnelling rates onto the dot from either spin up or spin
down leads and $\gamma_{\uparrow/\downarrow}^{>}$ are the tunnelling
rates off the dot for different spin populations. They are given by\[
\gamma_{\uparrow/\downarrow}^{<}=\gamma_{L\uparrow/\downarrow}^{<}+\gamma_{R\uparrow/\downarrow}^{<},\quad\gamma_{\uparrow/\downarrow}^{>}=\gamma_{L\uparrow/\downarrow}^{>}+\gamma_{R\uparrow/\downarrow}^{>}\]
where the tunnelling rates from the left and right leads are given
by \begin{eqnarray*}
\gamma_{L\uparrow}^{<}=\gamma_{0}f(\mu_{L}-\hbar\omega_{0}), & \quad & \gamma_{L\uparrow}^{>}=\gamma_{0}\left[1-f(\mu_{L}-\hbar\omega_{0})\right]\\
\gamma_{L\downarrow}^{<}=\gamma_{0}f(\mu_{L}+\hbar\omega_{0}), & \quad & \gamma_{L\downarrow}^{>}=\gamma_{0}\left[1-f(\mu_{L}+\hbar\omega_{0})\right],\end{eqnarray*}
with similar equations for the right lead, $f(\epsilon)=1/(1+\exp(\beta\epsilon))$
being the Fermi function and the chemical potentials in the left and
right lead are given by $\mu_{L/R}=\mu-V_{g}\pm V_{sd}/2$ with $\mu$
being the chemical potential, $V_{g}$ the gate voltage and $V_{sd}$
the bias voltage. A similar rate equation holds for the population
of other energy eigenstate\begin{eqnarray}
\dot{p}_{\uparrow b} & = & \left(\left|b_{\downarrow}\right|^{2}\gamma_{\downarrow}^{<}+\left|b_{\uparrow}\right|^{2}\gamma_{\uparrow}^{<}\right)p_{\uparrow0}\nonumber \\
 &  & -\left(\left|b_{\downarrow}\right|^{2}\gamma_{\downarrow}^{>}+\left|b_{\uparrow}\right|^{2}\gamma_{\uparrow}^{>}\right)p_{\uparrow b}.\label{pb}\end{eqnarray}
Additionally the individual probabilities have to add up to the probability
to find the stationary spin in a spin up state $p_{\uparrow}=p_{\uparrow a}+p_{\uparrow b}+p_{\uparrow0}.$
The equivalent rate equations can be derived for the stationary spin
down subspace and $p_{\uparrow}+p_{\downarrow}=1.$ The stationary
state can be obtained for each stationary spin subspace from $\dot{p}_{\uparrow a/b/0}=0$
and $\dot{p}_{\downarrow a/b/0}=0$ and we obtain the average population
on the quantum dot as a function of the applied magnetic field, \begin{equation}
\left\langle n\right\rangle =1-p_{\uparrow}L(\Delta_{0\uparrow})-p_{\downarrow}L(\Delta_{0\downarrow})\label{avn}\end{equation}
with \begin{equation}
L(\Delta)=\frac{\Delta^{2}r_{\infty}+\Omega^{2}r_{0}}{\Delta^{2}+\alpha^{2}\Omega^{2}},\label{L}\end{equation}
The coefficients in the Lorentzian Eq.~(\ref{L}) are given by\[
r_{\infty}=\frac{\left(1-f_{+}\right)^{2}-f_{-}^{2}}{1+f_{-}^{2}-f_{+}^{2}},\quad r_{0}=\frac{\left(1-f_{+}\right)^{2}}{1+f_{-}^{2}-f_{+}^{2}},\]
and

\[
\alpha^{2}=\frac{1-f_{+}^{2}}{1+f_{-}^{2}-f_{+}^{2}}\]
with $f_{+}=\frac{1}{2}\left[f_{\uparrow L}+f_{\uparrow R}+f_{\downarrow L}+f_{\downarrow R}\right]$,
$f_{-}=\frac{1}{2}\left[f_{\uparrow L}+f_{\uparrow R}-f_{\downarrow L}-f_{\downarrow R}\right]$
and $f_{\uparrow/\downarrow x}=f(\mu_{x}\mp\hbar\omega_{0}),\quad x=L,R.$

\begin{figure}
\begin{centering}
\includegraphics[width=1\columnwidth]{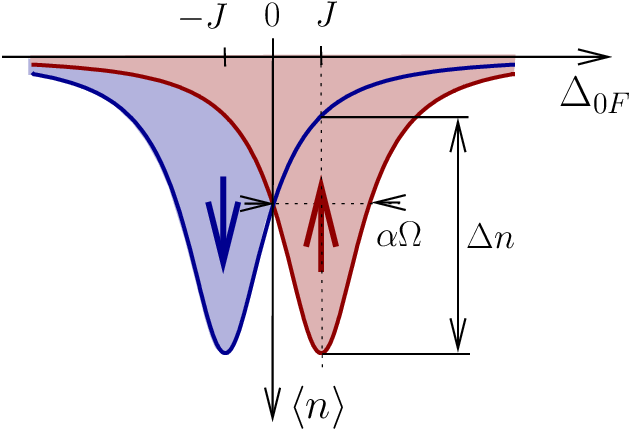}
\par\end{centering}

\caption{\label{fig4}The average dot population $\langle n\rangle$ as a function
of the detuning $\Delta_{0F}$ for stationary spin up and spin down
respectively. The two Lorentzian peaks are centred at $\pm J$ and
have a width of $\alpha\Omega$. Observing the population at a detuning
of $+J$ the difference in population between the case where the stationary
spin points up and where the stationary spin points down is given
by $\Delta n$. }

\end{figure}

The behaviour of the average dot population, Eq.~(\ref{avn}), as
a function of the detuning is shown in Fig. \ref{fig4}.%
 If we want to distinguish spin up from spin down we can monitor the
population at resonance, say $\Delta_{0F}=J$ and consider the change
in population from a pure spin up state ($p_{\uparrow}=1$) to a pure
spin down state ($p_{\downarrow}=1$)\[
\Delta n=L(0)-L(2J).\]

\begin{figure}
\begin{centering}
\includegraphics[width=1\columnwidth]{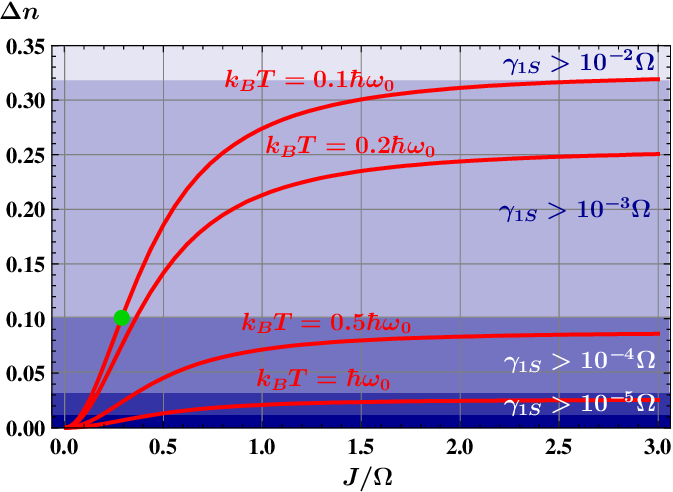}
\par\end{centering}

\caption{\label{fig5}Change of the charge on the quantum dot between spin
up and spin down on the stationary spin as a function of the exchange
coupling between the stationary and the flying spin for small bias
voltages ($V_{sd}=0.1\Omega$, $V_{g}=0$). Different curves are for
different temperatures. The blue shading in the background indicates
the detectable change in occupation given a certain relaxation time
of the stationary spin. The green dot indicates the example discussed
in the text.}

\end{figure}

The dependence of $\Delta n$ on the spin-spin interaction strength
$J$ and the temperature is shown in Fig.~\ref{fig5}.%
 The change in occupation number decreases for increasing temperature
and decreasing spin-spin interaction strength. In order to detect
a small change in the dot occupation number a certain number of electron
tunnelling events have to take place. To detect a change of $\Delta n$
more than $1/\Delta n^{2}$ electron have to tunnel, therefore the
electron tunnelling rate needs to be much larger than the stationary
spin relaxation time. In order to see a resonance of the spin on the
quantum dot two things have to be fulfilled: The Rabi frequency has
to be stronger than any spin relaxation, intrinsic or tunnelling induced,
and the thermal spin polarisation has to be significantly different
from zero. For a non-demolition measurement the difference in the
spin Zeeman splittings between stationary and flying spin has to be
much larger than the Rabi frequency and the spin-spin interaction. 

How does this translate to experimental parameters? With a spin coherence
time of $1\,\mu\textnormal{s}$ on the carbon nanotube quantum dot
we need a Rabi frequency of at least $\Omega=10\,\textnormal{MHz}$
to drive the spin into saturation, and thus effect a change of the
on dot spin occupation. Given the standard ESR microwave frequency
of $\omega_{0}=10\,\textnormal{GHz}$ and the corresponding magnetic
field of $350\,\textnormal{mT}$, at dilution refrigerator temperature
($T=50\,\textnormal{mK}$) we can read out a spin with a relaxation
time of $0.1\,\textnormal{ms}$ and a coupling of $J=3\,\textnormal{MHz}$.
To achieve such a coupling strength purely by dipole-dipole interaction
the spin must not be further than $1\,\textnormal{nm}$ from the nanotube
and the length of the dot must be $\approx100\,\textnormal{nm}$.
Such a setup would for example allow the readout of a molecular magnet
spin \citep{2007PhRvL..98e7201A}. Using a conductive molecule to
attach the spin to the nanotube could increase the coupling strength
or allow larger distances between nanotube and spin.

Quantum dots in nanotubes have been fabricated (see e.g. \citep{1998Natur.393...49T,2006ApPhL..89w2113J}),
also with integrated charge counters \citep{2006PhRvB..73t1402B,2008arXiv0810.0089G},
and on-chip microwave resonators reaching Rabi frequencies of $\sim10\,\textnormal{MHz}$
have been demonstrated (e.g. \citep{2003RScI...74.2749M,2005JMagR.175..275N}).
We therefore conclude that our scheme is realisable with current technology.
\begin{acknowledgments}
We thank the QIPIRC (GR/S82176/01) for support. BWL thanks the Royal
Society for a University Research Fellowship. JW thanks The Wenner-Gren
Foundations for financial support. We thank J. H. Wesenberg and S.
C. Benjamin for discussions.
\end{acknowledgments}

\end{document}